\documentclass[usenatbib,letter]{mn2e}
\usepackage{txfonts,graphicx}
\usepackage{amssymb}

\newcommand{\mum}{\ensuremath{\mu\mbox{m}}}

\newcommand{\hii}{H\,{\sc ii}}
\newcommand{\neii}{Ne\,{\sc ii}}
\newcommand{\neiii}{Ne\,{\sc iii}}
\newcommand{\nev}{Ne\,{\sc v}}
\newcommand{\siii}{S\,{\sc iii}}
\newcommand{\siv}{S\,{\sc iv}}
\catcode`\@=11
\def\gapprox{\mathrel{\mathpalette\@versim>}}
\def\lapprox{\mathrel{\mathpalette\@versim<}}
\def\@versim#1#2{\lower2.45pt\vbox{\baselineskip0pt\lineskip0.9pt
      \ialign{$\m@th#1\hfil##\hfil$\crcr#2\crcr\sim\crcr}}}
\catcode`\@=12
\newcommand{\araa}{ARA\&A}%
\newcommand{\apj}{ApJ}%
\newcommand{\apjs}{ApJS}%
\newcommand{\aap}{A\&A}%
\newcommand{\mnras}{MNRAS}%
\begin{document}

\title[The MIR {[\siv]/[\neii]} vs. {[\neiii]/[\neii]}
Correlation]{The Mid-Infrared [\siv]/[\neii] versus [\neiii]/[\neii]
Correlation}

\author[B.~Groves, B.~Nefs, \& B.~Brandl]{
     Brent Groves\thanks{brent@strw.leidenuniv.nl}, Bas Nefs, \& Bernhard Brandl\\
     Sterrewacht Leiden, Leiden University, Niels Bohrweg 2, 
     NL-2333 CA Leiden, The Netherlands} 
\date{Received <date> / Accepted <date>}
   
\maketitle
 
\begin{abstract}
   {The mid-infrared ratio [\neiii]$_{15.6\mu\rm{m}}$/[\neii]$_{12.8\mu\rm{m}}$
   is a strong diagnostic of the ionization state of emission line
   objects, due to its use of only strong neon
   emission lines only weakly affected by extinction. However this ratio
   is not available to ground-based telescopes as only a few
   spectroscopic windows are available in the MIR.} 
   {To deal with this problem we aimed to verify if there exists a
   conversion law between ground-accessible, strong MIR line ratio 
   [\siv]/[\neii] and the diagnostic [\neiii]/[\neii] ratio that can
   serve as a reference for future ground-based observations.} 
   {We collated the [\siv]$_{10.5\mu\rm{m}}$, [\neii]$_{12.8\mu\rm{m}}$,
   [\neiii]$_{15.6\mu\rm{m}}$ and [\siii]$_{18.7\mu\rm{m}}$ emission
   line fluxes from a wide range of sources in the rich \emph{Spitzer} and ISO
   archives, and compared the [\neiii]/[\neii], [\siv]/[\siii], and
   [\siv]/[\neii] ratios.} 
   {We find a strong correlation between the [\siv]/[\neii] and
   [\neiii]/[\neii] ratio, with a linear fit of log([\neiii]/[\neii]) $=
   0.81\log$([\siv]/[\neii]) $+0.36$, accurate to a factor of $\sim$2
   over four orders of magnitude in the line ratios. This demonstrates
   clearly the ability of ground-based infrared spectrographs to do
   ionization studies of nebulae.}   
\end{abstract}

\section{Introduction}
The mid-infrared spectrum (MIR, $\sim$3-40$\mu$m) hosts a range of
important diagnostic fine-structure emission lines.
Due to the much lower dust opacity at these long wavelengths,  the
emission lines are significantly less affected by extinction than
ultraviolet or optical lines \citep[see e.g.][]{Draine2003}. 
Thus these lines are able to probe 
ionized regions that lie behind dense obscuring clouds or even deep within
galaxy centers, giving insights into the regions where massive star
formation events actually occur, or heavily obscured
active galactic nuclei (AGN) can exist.

One of the more useful facets of this wavelength regime
is that it includes emission lines from several ions of the same
species, in particular strong lines from neon
and sulphur. These lines can be used as strong diagnostics of the ionization
state of the emitting gas as, being from the same species, there is no
direct dependence on the gas phase abundances, and the different
ionization potentials allow the observer to trace the hardness of the
ionizing spectrum \citep[see e.g.][]{Dopita2003}.
Lines ratios such as
[\nev]$_{14.3\mu\rm{m}}$/[\neii]$_{12.8\mu\rm{m}}$, 
[\neiii]$_{15.6\mu\rm{m}}$/[\neii]$_{12.8\mu\rm{m}}$, and
[\siv]$_{10.5\mu\rm{m}}$/[\siii]$_{18.7\mu\rm{m}}$ have been
used to diagnose the presence of deeply buried AGN
\citep[e.g.][]{Sturm2002,Armus2006} and as indicators of 
the ages of starbursts \citep[e.g.][]{Achtermann1995,Thornley2000}.

The major drawback of the MIR emission lines as diagnostics is their
accessibility, with the full range of lines only available to spectrographs
aboard space-borne telescopes like the \emph{short wavelength
spectrograph} \citep[SWS,][]{deGraauw1996} aboard the \emph{Infrared Space
Observatory} \citep[ISO,][]{Kessler1996} and the \emph{infrared spectrograph}
\citep[IRS,][]{Houck2004} aboard the \emph{Spitzer Space Telescope}
\citep{Werner2004}. While MIR instruments on ground based telescopes
can provide higher spatial and spectral resolution, only a sparse
number of MIR bands are available due to 
the opaqueness of the Earth's atmosphere in the mid-infrared 
(M-band:$\sim 4.5$ - 5\mum,
N-band:$\sim 8.5$ - 13\mum, and Q-band:$\sim 17$ - 23\mum). As a
result, the [\neiii]$_{15.6\mu\rm{m}}$ line is inaccessible from the
ground and therefore recent 
high-resolution studies have resorted to the ratio
[\siv]$_{10.5\mu\rm{m}}$/[\neii]$_{12.8\mu\rm{m}}$ as a diagnostic
\citep[e.g.][]{Snijders2007}, in a similar manner to earlier
aircraft-borne IR telescope studies \citep[e.g.][]{Achtermann1995}. However the
question remains whether the 
use of this ratio as a proxy for the ionization sensitive
[\neiii]$_{15.6\mu\rm{m}}$/[\neii]$_{12.8\mu\rm{m}}$ ratio is justified. While
photoionization theory clearly indicates that there will be a
correlation between these ratios, the dependence of the line emission
on often unknown physical parameters make the accuracy of this
correlation uncertain. As an example, the [\siv]/[\neii] ratio is
directly proportional to the sulfur to neon abundance ratio, while the
[\neiii]/[\neii] will be almost unaffected by any changes in the relative
abundances. Similarly, with an excitation potential of 34.79 eV,
[\siii] is sensitive to lower energy photons than [\neiii], which has
an excitation potential of 40.96eV and will not exist in photoionized
gas when there are no photons above this limit.

In this letter we use archival ISO and
\emph{Spitzer} observations of a range of astrophysical objects to
demonstrate that  such a correlation does exist and derive a simple
conversion law for these ratios. We mention possible systematics with
the determination of these ratios and discuss the origin of the
observed correlation. As a final note we
recommend the application of these results in future
(ground-based) observations.

\section{Observational data}\label{sec:data}
The strong, diagnostic, MIR emission lines [\siv]$_{10.5\mu\rm{m}}$,
[\neii]$_{12.8\mu\rm{m}}$, 
[\neiii]$_{15.6\mu\rm{m}}$, and [\siii]$_{18.7\mu\rm{m}}$  
have been detected and characterised in a broad range of astrophysical
objects, from nearby galactic \hii\ regions to ultraluminous infrared galaxies
(ULIRGs). To determine the connection between the [\siv]/[\neii] and
[\neiii]/[\neii] ratios we have extracted from the recent literature
the fluxes of these four emission lines. While not exhaustive, the sample of
355 emission line objects is
representative of the range of sources from which these lines
arise, covering a range of physical
parameters, such as source morphology and geometry, nature of the
ionizing sources and metal and dust abundances.

\begin{table}
\caption{Publications from which we sourced our MIR emission lines
differentiated the object classes, along with the telescope used and
the mean fractional
error in the [\neii]$_{12.8\mu\rm{m}}$ line. \label{tab:sources}}
\begin{tabular}{|lcl|}
\hline
Objects & $\sigma_{\rm{NeII}}/$[\neii] & Publication \\
\hline
Galactic \hii\ regions & 0.30                 & \citet{Giveon2002}$^2$       \\
Giant \hii\ regions & 0.15     & \citet{Lebouteiller2008}$^1$ \\
Extragalactic   &~& ~\\
~~\hii\ regions \& nuclei & 0.05    & \citet{Dale2006}$^1$         \\
M101 \hii\ regions & 0.39      & \citet{Gordon2008}$^1$       \\
M83 \hii\ regions & 0.01       & \citet{Rubin2007}$^1$    \\
M33 \hii\ regions & 0.01       & \citet{Rubin2008}$^1$        \\
M82 Pointings & 0.15           & \citet{Beirao2008}$^1$      \\
Arp244 Pointings & 0.04        & \citet{Brandl2008}$^1$       \\
Starburst galaxies & 0.09      & \citet{Brandl2006}$^{\ast,1}$ \\
Starburst galaxies & 0.30      & \citet{Verma2003}$^2$        \\
BCDs & 0.04                    & \citet{Wu2008}$^1$          \\
BCDs \& Starbursts & 0.09      & \citet{Engelbracht2008}$^1$  \\
ULIRGs & 0.10                  & \citet{Farrah2007}$^1$       \\
AGN & 0.42                     & \citet{Sturm2002}$^2$        \\
AGN  & 0.14          & \citet{Deo2007}$^1$          \\
LMC \& SMC PNe & 0.11          & \citet{Bernard-Salas2008}$^1$ \\
\hline
\end{tabular}\\
1 : \emph{Spitzer}\\
2 : \emph{ISO-SWS}
$\ast$ : Private Communication\\
\end{table}

A list of the different source types in our sample and the relevant
papers from which we extracted these is given in
Table \ref{tab:sources}. 
Included in the sample are sources ionized by massive stars, both
small scale (\hii\ regions and extragalactic \hii\ regions) and large
scale (blue compact dwarfs (BCDs), starburst galaxies (SB), and Ultra
Luminous IR galaxies (ULIRGs)), planetary nebulae (PNe), which are
ionized by white dwarfs, and active galactic nuclei (AGN). 

The table also includes 
the average percentage uncertainty in the [\neii] line as given in
the reference. This uncertainty varies greatly between the various
references, as the different authors have used different methods to
account for both the observational and systematic uncertainties for
their sources, which vary in flux and exposure times. While these
uncertainties may bias our fit, the close correlation observed and the
varying  sizes of the samples (both discussed in the next
section) will tend to minimize these effects. The given line
uncertainties in the references were
carried over when calculating their corresponding ratios. 

The majority of objects within our sample were observed with the
\emph{Spitzer} IRS, with the remainder coming from ISO SWS
observations (as noted in Table \ref{tab:sources}). In the cases
where objects were observed twice (for example with both
\emph{Spitzer} and ISO), we have included both observations, to
indicate the systematic uncertainty in the sample.
For all of our sample, all four MIR lines fall within \emph{Spitzer}'s IRS
short-high (SH) module (9.9-19.6$\mu$m) and ISO's SWS (2.38 to
45.2\mum), meaning that aperture effects are limited to that due to
the different wavelengths of the lines in the ratios (a factor of 1.2
for [\siv] to [\neii] and [\neii] to [\neiii]). 

Within our total sample there are approximately 8 objects observed by
both telescopes, and a similar number of objects observed by \emph{Spitzer} 
whose lines have been extracted by two different authors using
different methods. For objects by both telescopes (e.g.~NGC 4151), the
difference in line ratios is of the order of 0.3 dex, most likely due
to the different telescope apertures and pointings. Some objects were
observed once by ISO, but have multiple pointings by
\emph{Spitzer}-IRS (e.g.~M82), and for these objects we consider the
ISO observation an integrated flux and the IRS observations to be
individual \hii\ regions. For the few IRS observed objects
investigated by different authors, the differences in line ratios
arise purely due to the different 
line flux extraction techniques (e.g.~IIZw40)  
and are of the order of 0.2 dex, and within the given uncertainties.
However the small 
number of objects observed multiply ($\sim$15, distributed across both AGN
and Starburst/BCD) compared to the total sample size (355), mean that
a bias of the final fit towards these types of objects will not occur. 

In extracting our sample we excluded all objects 
in which any of the four lines were undetected, or for which
only upper limits were provided. This means that we do have a bias in
the sample against weak line objects, but, with objects covering
a broad range of distances and four orders of magnitude in line
ratio, this bias is likely to be weak. This bias is present, however,
in at least one of the groups (ULIRGs), and is discussed further in
the next section. 

The total sample represents a large variety in ionising conditions
($-2.1 < $log([\neiii]/[\neii])$ < 1.8$), metal abundances (ranging
from $\log (\rm{O}/\rm{H})+12 \approx 7.1$ (SBS 0335--052W) to
$\approx 8.8$ (IC342)) and the object particulars such as
infrared luminosities (PNe to ULIRGs), galaxy types 
(blue compact dwarf galaxies to  massive ULIRGs), and star formation
rates (BCDs and \hii\ regions of less than a few $M_{\odot}$ yr$^{-1}$
to ULIRGs $\sim 100$'s $M_{\odot}$ yr$^{-1}$).  

\section{Results \& Discussion}
\begin{figure}
\centering
\includegraphics[width=1.1\hsize]{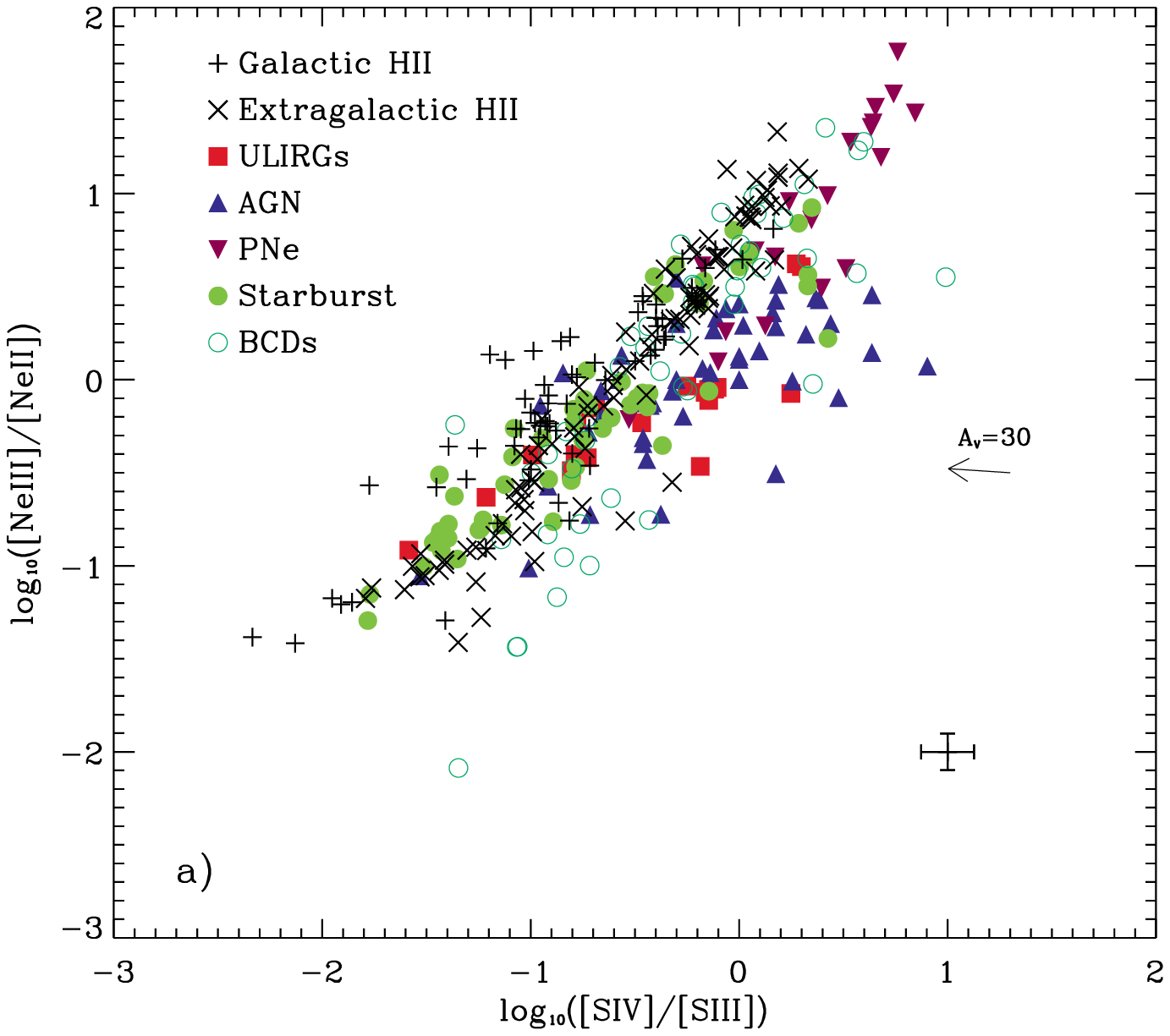}
\includegraphics[width=1.1\hsize]{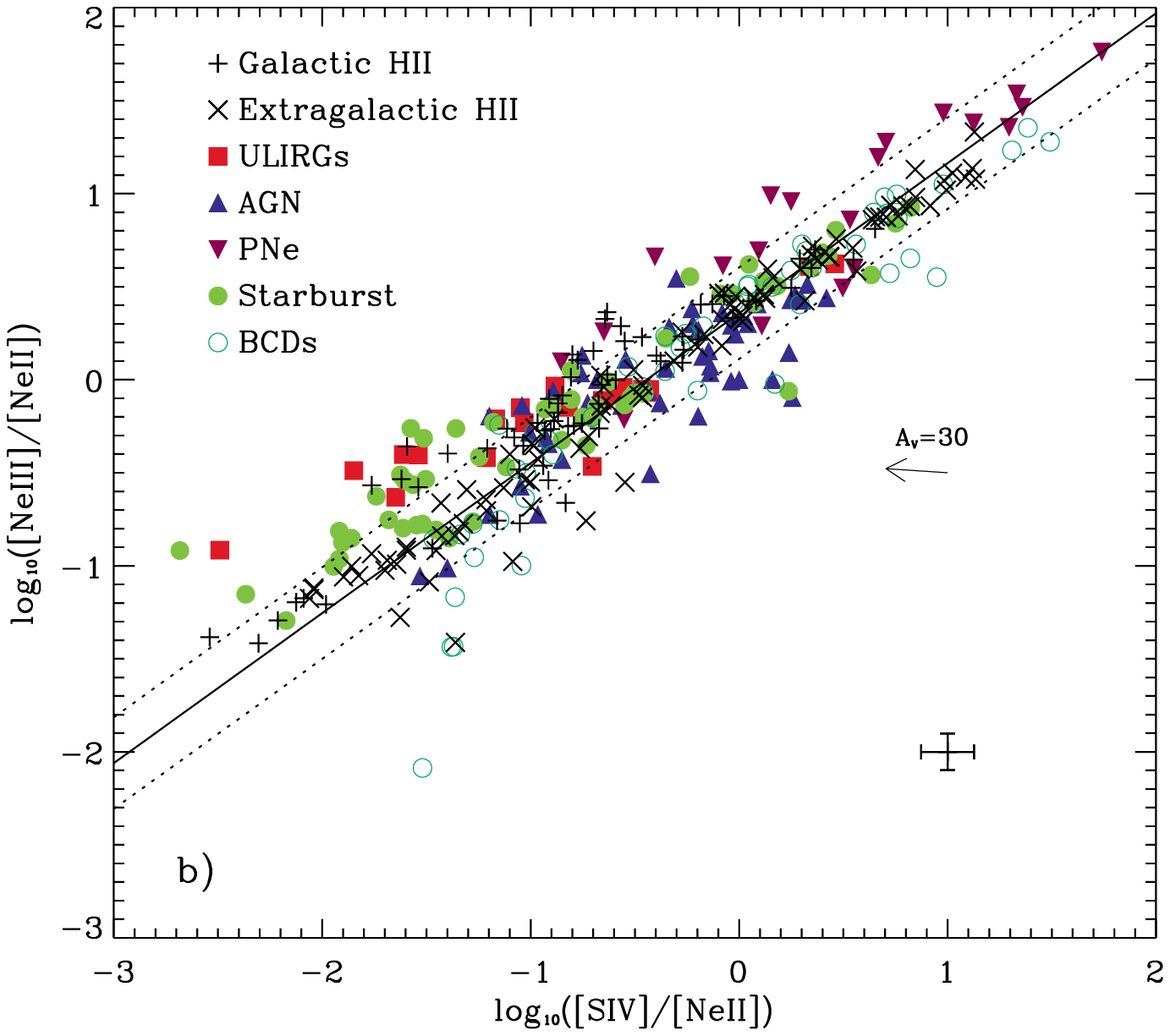}
\caption{\label{fig:obsdata} Mid-IR line ratio diagrams revealing the
correlation between [\neiii]$_{15.6\mu\rm{m}}$/[\neii]$_{12.8\mu\rm{m}}$ and
[\siv]$_{10.5\mu\rm{m}}$/[\siii]$_{18.7\mu\rm{m}}$ (a), and
[\neiii]$_{15.6\mu\rm{m}}$/[\neii]$_{12.8\mu\rm{m}}$ and
[\siv]$_{10.5\mu\rm{m}}$/[\neii]$_{12.8\mu\rm{m}}$ (b). Each diagram contains
the observed ratios from several different emission-line objects,
differentiated by colour and shape, as labelled by the key in the
upper left hand corner, and discussed in section \ref{sec:data}. The
error bars in the lower right of the diagrams indicate the average
uncertainty for the whole sample. The arrow indicates the effect of
extinction on the ratios for an A$_{\rm{V}}=30$ magnitudes.
The lower diagram also shows our best fit to the ratios (solid line)
along with the 1-$\sigma$ dispersion (dotted lines), as given in Table
\ref{tab:fits}.}
\end{figure}

The final sample of 355 emission-line objects is presented in figure
\ref{fig:obsdata} in the form of two emission line diagnostic
diagrams: [\siv]/[\siii] vs
[\neiii]/[\neii] (figure \ref{fig:obsdata}a) and [\siv]/[\neii] vs
[\neiii]/[\neii] (figure \ref{fig:obsdata}b).  
The separate classes of object within the sample
are differentiated both by colour and shape, as labelled in the
key in the upper left hand corner of the diagrams. In the lower right
corner of the diagrams we show the average observational uncertainty across
the sample, as given within the references. As noted previously, 
the uncertainty given in the references
varies significantly between the 
different classes, ranging from $\sim$1\% for M33 \& M83 to $\sim$40\% for
the galactic \hii\ regions. 
In all classes however, the uncertainty on the
[\neiii]/[\neii] ratio is lower than that of the other two ratios
considered here,
partly due to the strength of the neon lines (higher signal-to-noise), 
but mostly due to the
location of the [\siv]$_{10.5\mu\rm{m}}$ in the silicate absorption
feature (discussed further below), which both decreases the strength
of this line 
and adds an additional uncertainty due to extinction.
In any case, these uncertainties are applied to the reduced $\chi^2$
fit discussed below.

In both diagrams we plot a vector
that indicates the effects of an $A_{\rm V}=30$
magnitudes of extinction on the line ratios by a uniform dust screen,
assuming a 
\citet{Li2001} extinction curve. Note that, depending on which
curve is assumed, the length and direction of the extinction vector can
change but that in all cases, the [\siv]$_{10.5\mu\rm{m}}$ line suffers the
highest relative extinction 
\citep[see Table 2 and associated discussion in][]{Farrah2007}. 

The first diagram, figure \ref{fig:obsdata}a, reveals the standard
space-based diagnostic for the average ionization state of the emission line
object. This diagram is a strong diagnostic as it uses 4 strong lines,
with each  
ratio involving two different ionization states of the same element,
meaning that abundance differences are not an issue. Due to this
ionization dependence and 
insensitivity to abundance, this diagram has
been often used in the literature as a measure of the average hardness
of the ionizing radiation field, most recently by
\citet{Gordon2008}. In \citet{Gordon2008}, the correlation between the
[\siv]/[\siii] and [\neiii]/[\neii], clearly visible in our figure
\ref{fig:obsdata}a, was quantified for both individual \hii\ regions
in M101 and for a combined M101 and starburst/BCD galaxy sample from
\citet{Engelbracht2008} (their equations 5 and 6 respectively). They
find a steeper slope than appears in figure \ref{fig:obsdata}a (1.53
versus 1.29), but this may be due to the smaller sample size, and is
also biased by the \citet{Engelbracht2008} starburst sample. This sample,
marked as BCDs in our sample, can be seen as the offset empty circles
at low [\neiii]/[\neii] in figure \ref{fig:obsdata}a. The visible
offset from the rest of the sample may be intrinsic, but may also be
from the underestimation of the contribution from the 12.7\mum\ PAH
feature in these weaker line objects \citep{Smith2007}.

There are several offset groups within figure
\ref{fig:obsdata}a, the most noticeable being the AGN. This group is
clearly offset towards higher [\siv]/[\siii] values relative to their
[\neiii]/[\neii] ratio. The reason for this offset is uncertain, but
is most likely related to the harder, power-law  
ionizing spectrum found in AGN. This leads to higher average
electron temperatures in the S$^{++}$ and Ne$^{+}$ zones and different
extents, relative to gas ionized by stars, and therefore a weaker
[\siii]/[\neii] ratio for AGN relative to stellar photoionization.
This offset may also be 
measurement related, with the weaker 12.7\mum\ PAH feature meaning
that the [\neii] line is determined more accurately and with higher
flux. The presence of ULIRGs and some starburst galaxies in this
region as well is unsurprising as some of these sources contain known
active nuclei,  just as many of the AGN classified galaxies also have
strong, active star formation.

In figure \ref{fig:obsdata}b we show the ground-accessible ratio
[\siv]/[\neii] versus [\neiii]/[\neii], revealing a strong correlation
between the ratios, justifying the use of the [\siv]/[\neii] ratio as
a surrogate for the [\neiii]/[\neii] ratio when unavailable.
Compared with figure \ref{fig:obsdata}a, the data show surprisingly
less scatter. This probably arises due to the use of a common
denominator ([\neii]), but the [\siii] line
may also be introducing some scatter into figure \ref{fig:obsdata}a,
possibly due to its longer wavelength (and hence lower resolution).

To quantify this correlation we perform a linear least-squares fit to
the full sample using the IDL routine
FITEXY, giving the empirical relationship,
\begin{equation}
\log\left([\rm{NeIII}]/[\rm{NeII}]\right) =
\alpha\log\left([\rm{SIV}]/[\rm{NeII}]\right)+\beta. 
\end{equation} 
The result of this fit, along with the 1 $\sigma$ dispersion, is
given in Table \ref{tab:fits}. 

\begin{table}
\caption{Fit parameters for the full sample and individual classes,
including the  number of objects (N) within the sample. \label{tab:fits}}
\centering
\begin{tabular}{lcccc}
\hline
Object class & N & $\alpha$ & $\beta$ & $\sigma_{Y|X}$ \\
\hline
Full sample         &  355 &   0.81 &   0.36 &  0.25 \\
\hii\ regions       &   65 &   0.74 &   0.42 &  0.23 \\ 
Extragalactic \hii  &   97 &   0.82 &   0.36 &  0.15 \\ 
ULIRGs              &   20 &   0.49 &   0.26 &  0.33 \\
AGN                 &   20 &   0.69 &   0.29 &  0.25 \\
PNe                 &   20 &   0.74 &   0.56 &  0.30 \\
Starburst           &   56 &   0.65 &   0.32 &  0.28 \\
BCDs                &   49 &   0.86 &   0.32 &  0.30 \\
\hline
\end{tabular}
\end{table}

The first thing to note is that the log-space linear fit to the full sample
appears to be accurate to 0.25 dex (that is, less than a factor of
$\sim$2 in the linear ratio) over four orders of
magnitude in the line ratios, demonstrating clearly the ability of
ground-based IR spectrographs to do ionization studies of nebulae.
The dispersion around this relation is greater than the average
uncertainty, indicating either inaccurate uncertainties, intrinsic
variations due to the variety of sources or, most likely, a
combination of both. 

In addition to the full fit we also present fits to our various
object classes, also presented in Table \ref{tab:fits}. Overall,
the relations match reasonably well with the full sample fit, with the
two major outliers either side being the planetary nebulae (high
offset in $\beta$) and the
ULIRGs (shallow slope in $\alpha$). 
These individual fits can be used when the specific class of
objects being observed are known, but we suggest that the full fit
should be used in most circumstances, as it is a more robust fit and
includes all object classes.

In figure \ref{fig:obsdata}b the PNe are visibly offset from the rest
of the sample, both to high [\siv]/[\neii] and higher than average
[\neiii]/[\neii]. This offset is most likely due to the hotter and harder
white dwarf ionizing source in PNe, although this then poses the
question why the AGN, which also have a harder ionizing spectrum than that
found in \hii\ regions and galaxies, is not also more offset.

The shallow slope of the ULIRG sample demonstrates one of the 
issues with the [\siv]/[\neii] ratio: the
location of the [\siv]$_{10.5\mu\rm{m}}$ in the $9.7\mum$ silicate
absorption feature. Of the \citet{Farrah2007} sample of 53 ULIRGs over
half have no detections or upper limits for the [\siv] line, while a
large fraction of 
those which do have detections are uncertain, with detections of less
than 3 sigma. 
As noted in the paper, only in ULIRGs with low silicate depths
($S_{\rm sil} \lapprox 2.1$) is the [\siv] feature detected, even when
the [\neiii]$_{15.6\mu\rm{m}}$ line is present. Thus there is likely both a
bias in our ULIRG sample and an offset, such that low [\siv]/[\neii]
ULIRGs are also those with high extinction and therefore offset from
the relation, while the highest extinction sources are not included at all..

Another spectral feature which may affect the line ratio relation is
the 12.7\mum\ PAH feature seen clearly in many galaxy spectra
\citep{Smith2007}. This broad feature lies below the [\neii] line, and makes
the determination and subtraction of the underlying continuum
difficult, and thus increases the uncertainty in the flux of this
line. As mentioned before, this may explain the offset seen in the
\citet{Engelbracht2008} sample.

However, even with these outliers the correlation is surprisingly
tight. Simple photoionization theory expects such a correlation, with
higher ionization leading to higher values for both ratios
\citep{Dopita2003}, but  
variations in the ionization parameter\footnote{The ionization
parameter is a measure of the ionizing photon density over the gas
density; $U = Q_{\rm{tot}}/n_{\rm{H}}c$, where $Q_{\rm{tot}} =
\int_{13.6\rm{eV}} F_{\nu}/h\nu~d\nu$} and ionizing spectrum cause 
different changes in the two ratios. Such variations have been
explored in the case of starbursts and \hii\ regions in
\citet{Thornley2000} and \citet{Rubin2008} and, given these results,
the tight fit indicates a close correlation between ionization
parameter and spectral hardness in these objects.

One further issue is abundance variations. The [\neiii]/[\neii] ratio
is sensitive to abundance variations only indirectly, through the
resulting temperature effects. The [\siv]/[\neii] ratio however, can be
directly affected by variations between the sulphur and neon
abundances. As both are
primary elements, significant total abundance variations are
not expected. However, depletion of sulphur onto dust may cause
variations in the gas phase and hence in the emission lines. While the total
depletion of sulphur onto dust is uncertain \citep[though
see][]{Lebouteiller2008}, it does not seem to be a 
great effect on the Ne/S ratio, as shown by
\citet{Rubin2007,Rubin2008}, and therefore on the [\siv]/[\neii]
ratio. 

Thus, in summary, to determine whether it is reasonable to use the
ground-accessible 
ratio [\siv]$_{10.5\mu\rm{m}}$/[\neii]$_{12.8\mu\rm{m}}$ as a
replacement  for 
the diagnostic [\neiii]/[\neii] ratio, we have collated existing ISO and
\emph{Spitzer} spectral observations of a wide range
of astrophysical objects from the literature. We find a good
correlation between these ratios with a linear fit giving the
relation,
\begin{displaymath}
\log\left([\rm{NeIII}]/[\rm{NeII}]\right) =
0.81\log\left([\rm{SIV}]/[\rm{NeII}]\right)+0.36, 
\end{displaymath} 
with a 1 $\sigma$ dispersion of 0.25, corresponding to an uncertainty in the
estimated line ratio of $\sim70$\%. We propose that for future
ground based observations this relationship be used to determine the
ionization state of the observed objects, though note that caution must
be applied when looking at heavily extinguished ($A_{\rm V} \gapprox
20$) objects where the [\siv]$_{10.5\mu\rm{m}}$ may be affected by the
9.7\mum\ silicate absorption feature.

\section*{Acknowledgments}
Part of this work was supported by the German
\emph{Deut\-sche For\-schungs\-ge\-mein\-schaft, DFG\/} project
number Ts~17/2--1.

\end{document}